\documentclass[aps,graphicx,twocolumn]{revtex4}

\usepackage{amsmath} 
\usepackage{mathrsfs}
\usepackage[cmintegrals]{newtxmath}
\usepackage{graphicx}
\usepackage{dcolumn}
\usepackage{hyperref}
\usepackage{bm}
\usepackage[mathlines]{lineno}

\begin{document}

\title{Multimode interference induced optical nonreciprocity and routing in an optical microcavity}

\author{Hong Yang$^{1,2}$, Guo-Qing Qin$^{1,2}$, Hao Zhang$^{1,2}$, Xuan Mao$^{1,2}$, Min Wang$^{1,2,4}$ and Gui-Lu Long$^{1,2,3,4,5}$\footnote{gllong@tsinghua.edu.cn}}

\address{$^1$State Key Laboratory of Low-Dimensional Quantum Physics, Department of Physics, Tsinghua University, Beijing 100084, China\\
$^2$Frontier Science Center for Quantum Information, Beijing 100084, China\\
$^3$Beijing National Research Center for Information Science and Technology, Beijing 100084, China\\
$^4$Beijing Academy of Quantum Information Sciences, Beijing 100193, China\\
$^5$School of Information, Tsinghua University, Beijing 100084, China
}

\date{\today}

\begin{abstract}
Optical nonreciprocity and routing using optocal microcavities draw much atttention in recent years. Here, we report the results of the study on the nonreciprocity and routing using optomechanical multimode interference in an optical microcavity. The optomechanical system used here possesses multi-optical modes and a mechanical mode. Optomechanical induced transparency and absorption, appear in the system due to the interference between different paths. The system can present significant nonreciprocity and routing properties when appropriate parameters of the system are set. We design quantum devices, such as diode, circulator and router, which are important applications. Our work shows that optomechanical multimode system can be used as a promising platform for buliding photonic and quantum network.
\end{abstract}

\maketitle

\section{Introduction}

High quality optical microcavities \cite{vahala2003optical}, which can enhance light–matter interactions in a very confined volume, play an essential role in optical physics study and applications. Examples of such applications include parity-time-symmetry \cite{peng2014parity,jing2014pt,chang2014parity}, chaos \cite{jiang2017chaos,lu2015p}, microcavity sensors \cite{sensing,chen2017exceptional,zhang2017far,ward2018nanoparticle,qin2019brillouin,wang2018rapid}, and etc. As a promising platform for realizing quantum electrodynamics, cavity optomechanics in optical microcavity have been widely investigated theoretically and experimentally \cite{aspelmeyer2014cavity,probing,arcizet2006radiation,Qin:20}. Early studies are restricted to basic optomechanical models with one optical mode and one mechanical mode, models with multimode interaction, which couples multiple optical modes to a mechanical mode, exhibit richer physics phenomena such as optomechanical induced transparency (OMIT) \cite{weis2010optomechanically,kronwald2013optomechanically,safavi2011electromagnetically,kim2015non,lu2018optomechanically,dong2012optomechanical,Long1,Long2} and absorption (OMIA) \cite{PhysRevA.71.043804,qu2013phonon}  and shows enormous potential in applications ranging from quantum information processing to state transfers \cite{kuzyk2017controlling,liao2016macroscopic,liu2013dynamic,stannigel2012optomechanical,zhang2019fast,liu2019sensing,wang2019chemo,chen2019phononic,liu2018optothermal,wang2012generation,wang2012quantum,wei2015hybrid,wang2020entanglement,wang2012using,tian2012adiabatic,xu2019frequency}. 

In recent years, as an intriguing physical phenomena, optical nonreciprocity has attracted great attentain, especially for nonreciprocity devices, which are widely used in diodes or isolators, the fundamental buliding-blocks of information network. Traditionally, optical nonreciprocity is associated with Faraday rotation effect \cite{Faraday} by breaking the time-reversal symmetry. Besides, nonmagnetic optical nonreciprocity, including nonlinear optics \cite{fan2012all}, optoacoustic effects \cite{kang2011reconfigurable} and parity-time-symmetric structures \cite{ruter2010observation}, can also be achieved. Optomechanical systems provide a promising platform for studying nonreciprocity \cite{dong2015brillouin,miri2017optical,manipatruni2009optical,bernier2017nonreciprocal,gui2006general,long2018realistic,qin2019proposal} induced by optomechanical multimode interference. Using this system, optical isolators \cite{shen2016experimental,ruesink2016nonreciprocity}, circulators \cite{ruesink2018optical,xu2015optical}, and directional amplifiers \cite{shen2018reconfigurable} have been theoretically studied and experimentally realized. Optical router is another key element for controlling the path of signal flow in quantum and classical network, it can be also constructed through muiltimode interference. Recently, quantum router has been proposed in various systems, i.e., cavity atom \cite{zhou2013quantum,shomroni2014all,yan2018targeted,li2015designable,lu2014single,aoki2009efficient,hoi2011demonstration,chen2013all,miller2010optical}, coupled resonator  \cite{xia2013all} and optomechanical systems \cite{agarwal2012optomechanical,fang2016optical}.

 In this paper, we propose a scheme to realize controllable optical nonreciprocity and  routing based on multimode interference in an optical cavity. The multimode system in our proposal is realized by coupling multi optical modes with a mechanical mode in an optical cavity. The multi optical modes are composed of a pair of coupled clockwise (CW) and counter-clockwise (CCW) modes. Fundamental physical phenomena, such as OMIT and OMIA can induced in the system. When signal input upon port 1 and port 2, which are in opposite directions, our system exhibits tunable nonreciprocity and the isolation ratio can be controlled by adjusting the optomechanical coupling strength. With this property, basic devices such as opitcal diodes and circulators can be designed. For transmission spectrum on three or four ports, the system presents controllable routing properties, which can be used for building quantum router. The result shows that photon can be transferred  from the input port to an arbitrarily selected output port with nearly 100$\%$ transmission.
 
 This article is organized as follows:  We describe the basic model of multimode interactions in optical microcavities in Sec.\ref{sec2}. In Sec.\ref{sec3}A, we study the multimode interference, and we show how to realize the tunable nonreciprocity and circulator in Sec.\ref{sec3b}. Three-port and four-port routing schme are given in Sec.\ref{sec4a} and B, respectively. Conclusion is given in Sec.\ref{sec5}.

\begin{figure}[!ht]
    \begin{center}
    \includegraphics[width=8.5cm,angle=0]{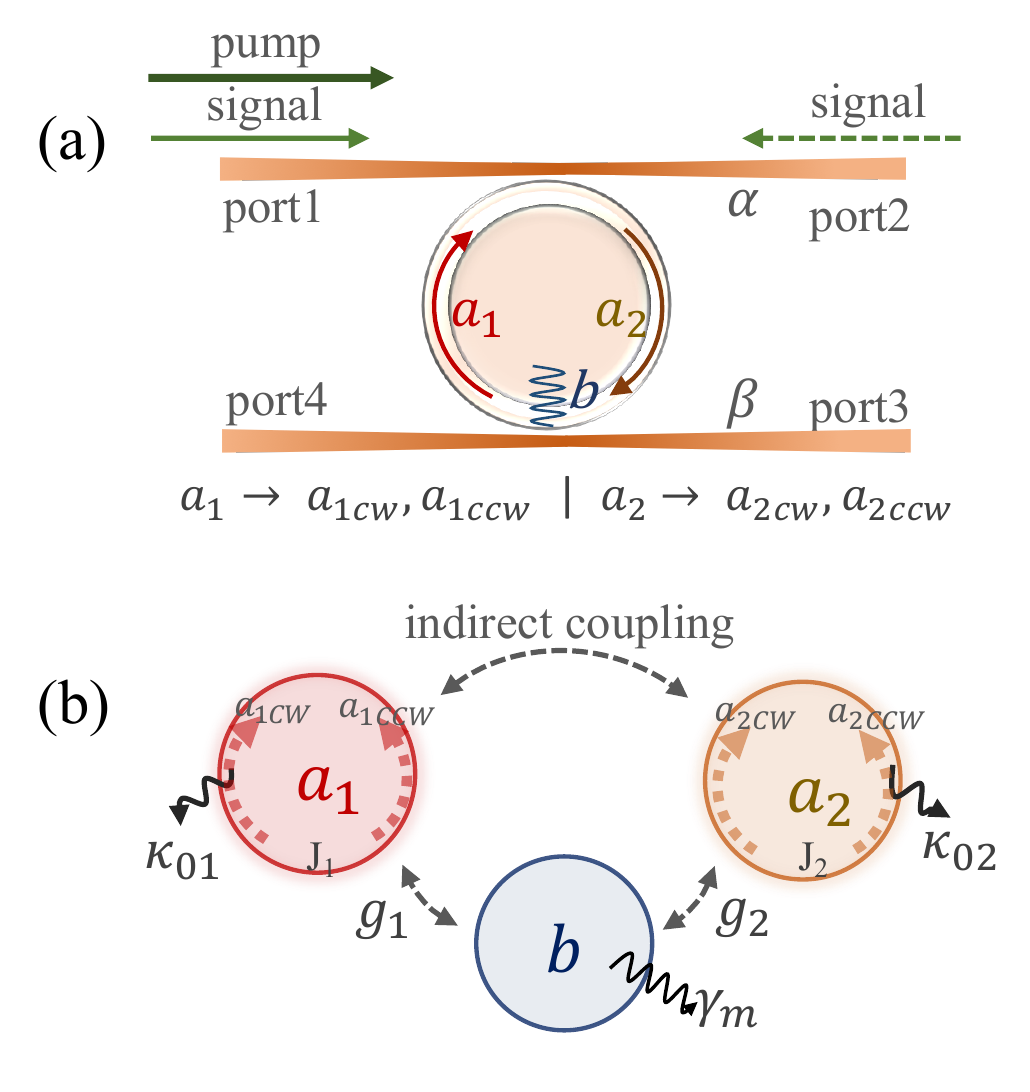}
    \caption{Schematic diagram of the four-port multimode optomechanical system. (a) Generic setup for optomechanical system: a whispering-gallery mode cavity mode is coupled with two fibers $\alpha$ and $\beta$. A field pumped into port 1 excites the coupling between the mechanical mode and the CW optical mode. (b) Principle description: two indirectly coupled  optical modes $a_{i}$ at frequency $\omega_{i}$ are coupled to a mechanical mode at frequency $\omega_m$ with coupling rates $g_1$ and $g_2$, respectively. Both optical modes support pairs of degenerate CW and CCW modes with coupling strength $J_{i}$} \label{basicmodel}
    \end{center} 
\end{figure}

\section{Basic model of multimode interaction} \label{sec2}
We consider a four-port optomechanical system shown in Fig.~\ref{basicmodel} (a), where a microcavity resonator is coupled with two fibers $\alpha$ and $\beta$ through evanescent field simultaneously. Two cavity modes $a_1$ and $a_2$, with corresponding frequencies $\omega_1$ and $\omega_2$, are coupled to the same mechanical mode $b$ with frequencies $\omega_m$ via radiation pressure. In addition, the whispering-gallery modes $a_1$ and $a_2$ couple indirectly through the fiber. Both the optical modes $a_1$ and $a_2$ support a pair of CW and CCW modes with oppsite travelling directions and with coupling strength $J_{1}$ and $J_2$, as shown in Fig.~\ref{basicmodel} (b). By pumping a laser field with frequency $\omega_p$ and amplitude $\varepsilon_{p}$ from port 1, which excites the coupling between the mechanical mode and the CW optical field in the cavity, the Hamiltonian of the system is given by ($\hbar=1$) 
\begin{align} 
    H = H_0+H_I+H_c+H_p \label{eq1}.
\end{align}
The first term in the right hand
\begin{align}
    H_0&=\sum\limits_{i,j}\omega_ia^\dagger_{i,j} a_{i,j}+\omega_mb^\dagger b \label{eq2}
\end{align}
describes the free Hamiltonian of each mode of the system, where $a_{i,j}$ ($a_{i,j}^{\dag}$) $(i=1,2;j=cw,ccw)$ denotes the annihilation (creation) operators of $i$-th CW and CCW modes, respectively, and $b$ ($b^{\dag}$) is the annihilation (creation) operator of the mechanical mode. The second term
\begin{align}
    H_I=\!\!\sum\limits_{i,j}g_ia^\dagger_{i,j} a_{i,j}(b^\dagger +b)+\!\!\sum\limits_j J_i(a^\dagger_{i,cw} a_{i,ccw}\!+a^\dagger_{i,ccw} a_{i,cw}) \label{eq3}
\end{align}
denotes the multimode interaction. The first term of $H_I$ represents the interactions between the cavity and mechanical mode, and the second one describes the coupling between the CW and CCW mode. $g_i$ is the single-photon optomechanical coupling rate. And $J_i$ is the photon-hopping strength between the CW and CCW mode due to optical backscattering. The third term
 \begin{align}
    H_c=&\sum_k\int_{-\infty}^{+\infty} \hbar \omega c_k^\dagger(\omega)c_k(\omega)d\omega\nonumber\\&+i \int_{-\infty}^{+\infty} \sum_{i,j,k} \kappa_{i,k}(\omega)[c_k^\dagger(\omega)a_{i,j}-a^\dagger_{i,j} c_k(\omega)]d\omega \label{eq4}
\end{align}
represents the Hamiltonian of the $\alpha$ and $\beta$ waveguide mode, the coupling between the optical modes and two waveguide modes, respectively. The annihilation operator $c_k(\omega)$ ($k=\alpha,\beta$) denotes the waveguide mode with commutation relation $[c_k(\omega),c_k^\dagger(\omega') ]=\delta(\omega-\omega')$. Parameter $\kappa_{i,k}(\omega)$ is the external loss rate between the cavity mode $a_{i,j}$ and the fiber $\alpha$ and $\beta$ . The last term
\begin{align}
    H_p&=i \sum_{i}\sqrt{\kappa_{i,\alpha}}\varepsilon_{p}(a^\dagger_{i,cw}e^{-i\omega_p t}-H.c.) \label{eq5}
\end{align}
describes the pump field. When a probe laser, with amplitude $\varepsilon_{s}^n$ and frequency $\omega_s$, is input into the system from port $n$ ($n=$ 1,2,3,4), considering dissipation and quantum (thermal) noise, the quantum Langevin equations for the operators in the rotating frame of the pump fields are given by
\begin{align}
    \frac{\mathrm{d}a_{i,cw}}{\mathrm{d}t}=&\;\Gamma_i a_{i,cw}-i J_i a_{i,ccw}-A a_{3-i,cw}\nonumber+\sqrt{\kappa_{i,\alpha}} \varepsilon_p\\&+(\sqrt{\kappa_{i,\alpha}} \varepsilon_{s}^1+\sqrt{\kappa_{i,\beta}} \varepsilon_{s}^3) e^{-i\delta t}+\sqrt{\kappa_{i,\alpha}}\xi_i,\label{eq6}\\
    \frac{\mathrm{d}a_{i,ccw}}{\mathrm{d}t}=&\;\Gamma_i a_{i,ccw}  - i J_i a_{i,cw}-A a_{3-i,ccw}\nonumber\\&+(\sqrt{\kappa_{i,\alpha}} \varepsilon_{s}^2+\sqrt{\kappa_{i,\beta}} \varepsilon_{s}^4) e^{-i\delta t}+\sqrt{\kappa_{i,\alpha}}\xi_i,\label{eq7}\\
    \frac{\mathrm{d}b}{\mathrm{d}t}=(-i&\omega_m\!-\!\frac{\gamma_m}{2})b-i\!\!\sum_i (g_i a^\dagger_{i,cw}a_{i,ccw}\!+\!H.c.)\!+\!\sqrt{\gamma_{m}}\xi_m,\label{eq8}
\end{align}
where the coefficients $\Gamma_i $ and $A$ are 
\begin{align}
    \Gamma_i = [i \Delta_i-\frac{\kappa_i}{2}-i g_i (b^\dagger+b)],\\
    A = \frac{\sqrt{\kappa_{1,\alpha}\kappa_{2,\alpha}}}{2}+\frac{\sqrt{\kappa_{1,\beta}\kappa_{2,\beta}}}{2}.
\end{align}
$A$ represents the interaction between optical mode and waveguide $\alpha$ and $\beta$, and $\Delta_i$  = $\omega_p-\omega_i$ denotes the detuning between the driving field and the cavity mode. 
 $\kappa_i$ is the total loss rate which contains an intrinsic loss rate $\kappa_{0i}$ and external loss rate $\kappa_{i,k}$ and $\delta = \omega_s-\omega_p $ is the detuning between the probe field and the pump field. Using standard linearization, the steady-state solution of the Eq. (\ref{eq6}) - (\ref{eq8}) in the red detuning can be obtained as follows:
 \begin{widetext}
 \begin{align}
    a_{1,cw}=\frac{(\tau_2 \sqrt{\kappa_{2,\alpha}}\!+\!\zeta_2 \sqrt{\kappa_{1,\alpha}})\varepsilon_{s}^1\!-\!i J_1(\tau_2 \alpha_2\!+\!\zeta_2\alpha_1)\varepsilon_{s}^2+(\tau_2 \sqrt{\kappa_{2,\beta}}\!+\!\zeta_2 \sqrt{\kappa_{1,\beta}})\varepsilon_{s}^3\!-\!i J_1(\tau_2 \beta_2\!+\!\zeta_2\beta_1)\varepsilon_{s}^4}{\zeta_1 \zeta_2 - \tau_1 \tau_2},\label{eq9}
\end{align}
\end{widetext}
\begin{align}
    a_{2,cw}=\frac{\tau_1a_{1,cw}+\sqrt{\kappa_{2,\alpha}}\varepsilon_{s}^{1}-\alpha_2\varepsilon_{s}^{2}+\sqrt{\kappa_{2,\beta}}\varepsilon_{s}^{3}-\beta_2\varepsilon_{s}^{4}}{\zeta_2},\label{eq10}
\end{align}
\begin{align}
    a_{1,ccw}=\alpha_1\varepsilon_{s}^2+\beta_1\varepsilon_{s}^4+i\mu(J_2Aa_{2.cw}-J_1\sigma_2a_{1,cw}),\label{eq11}
\end{align}
\begin{align}
    a_{2,ccw}=\alpha_2\varepsilon_{s}^2+\beta_2\varepsilon_{s}^4+i\mu(J_1Aa_{1.cw}-J_2\sigma_1a_{2,cw}),\label{eq12}
\end{align}

where the coefficients are
 \begin{align}
    &\zeta_i = \sigma_i+\frac{|G_i|^2}{\gamma_m}+\mu J_iJ_1\sigma_{3-i},\nonumber\\
    &\tau_i = \mu J_iJ_1A-\frac{{G_i}^*G_{3-i}}{\gamma_m}-A,\nonumber\\
    &\alpha_i = \mu (\sqrt{\kappa_{1,\alpha}}\kappa_{3-i}-A\sqrt{\kappa_{3-i,\alpha}}),\nonumber\\
    &\beta_i = \mu (\sqrt{\kappa_{1,\beta}}\kappa_{3-i}-A\sqrt{\kappa_{3-i,\beta}}),\nonumber\\
    &\mu =1/(\sigma_1\sigma_2-A^2), \nonumber \sigma_i = i(\omega_m - \delta)+\frac{\kappa_i}{2}\nonumber. 
\end{align} 
According to the input-output relation \cite{wallsquantum}, one can obtain the transmission spectrum of different ports.

\section{Multimode interference and Optical nonreciprocity}
\label{sec3}

\subsection{OMIT and OMIA in multimode interference}\label{sec3a}

\begin{figure}[!ht]
    \begin{center}
    \includegraphics[width=8.5cm,angle=0]{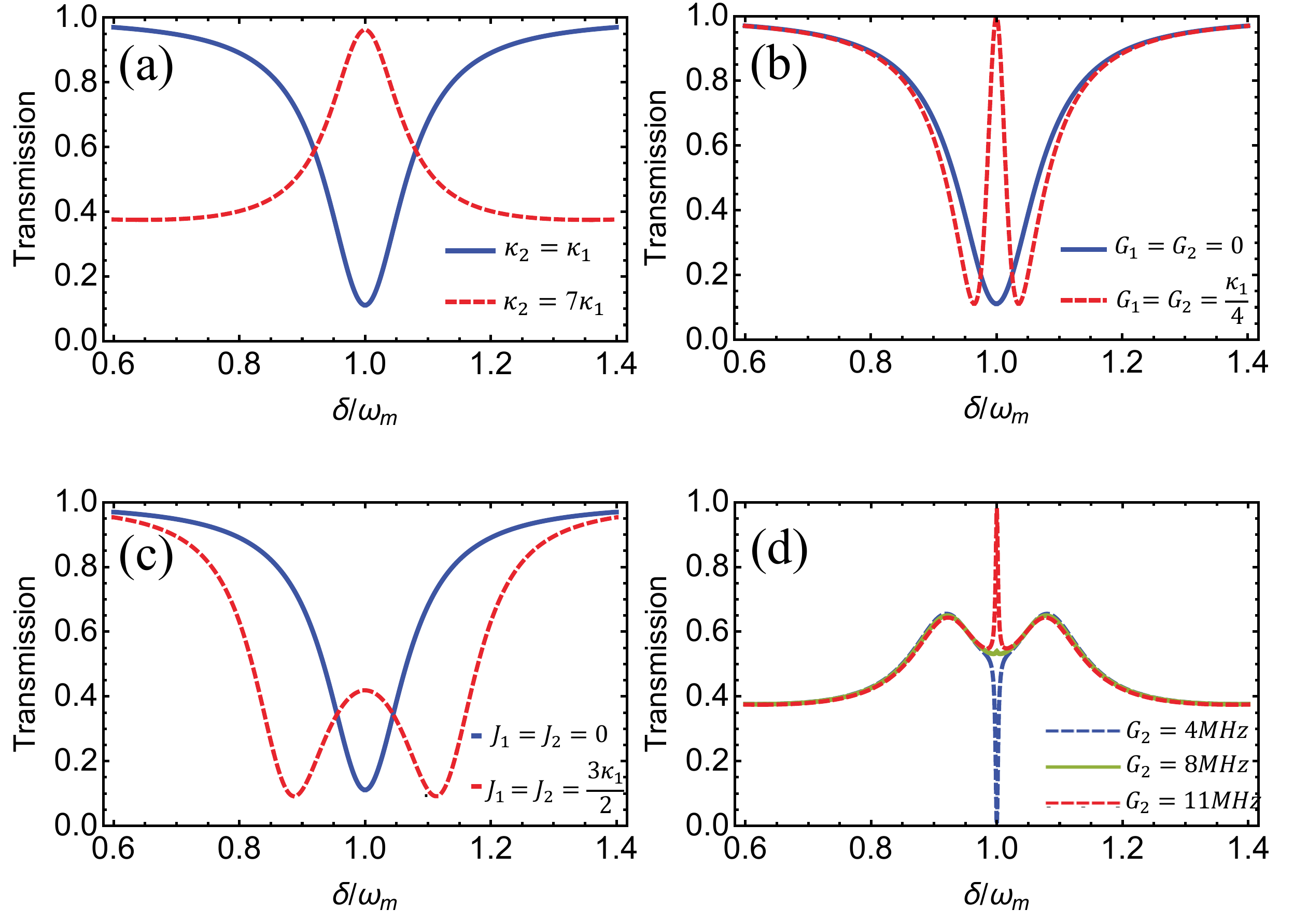}
    \caption{Transmission spectrum due to different interference ways. (a) Two optical modes couple indirectly through the waveguide inducing EIT-like line shape; (b) Pump field excites the coupling between the mechanical motion and the CW optical field, leading to OMIT phenomenon; (c) The CW and CCW modes are coupled with each other via scattering on the resonator surface, resulting in mode splitting; (d) All three interference co-exist, the system exhibits OMIT and OMIA. The parameters of red dashed line are (a) $G_1=G_2=$ 0, $J=$ 0. (b) $J$= 0, (c) $G_1=G_2=$ 0, (d)$G_1=$ 5 MHz, $J_1=J_2=$ 15 MHz. Other parameters are $\omega_m=$ 200 MHz, $\gamma_m=$ 5 KHz.}\label{interference} 
    \end{center} 
\end{figure}

\begin{figure*}[!ht]
    \begin{center}
    \includegraphics[width=16cm,angle=0]{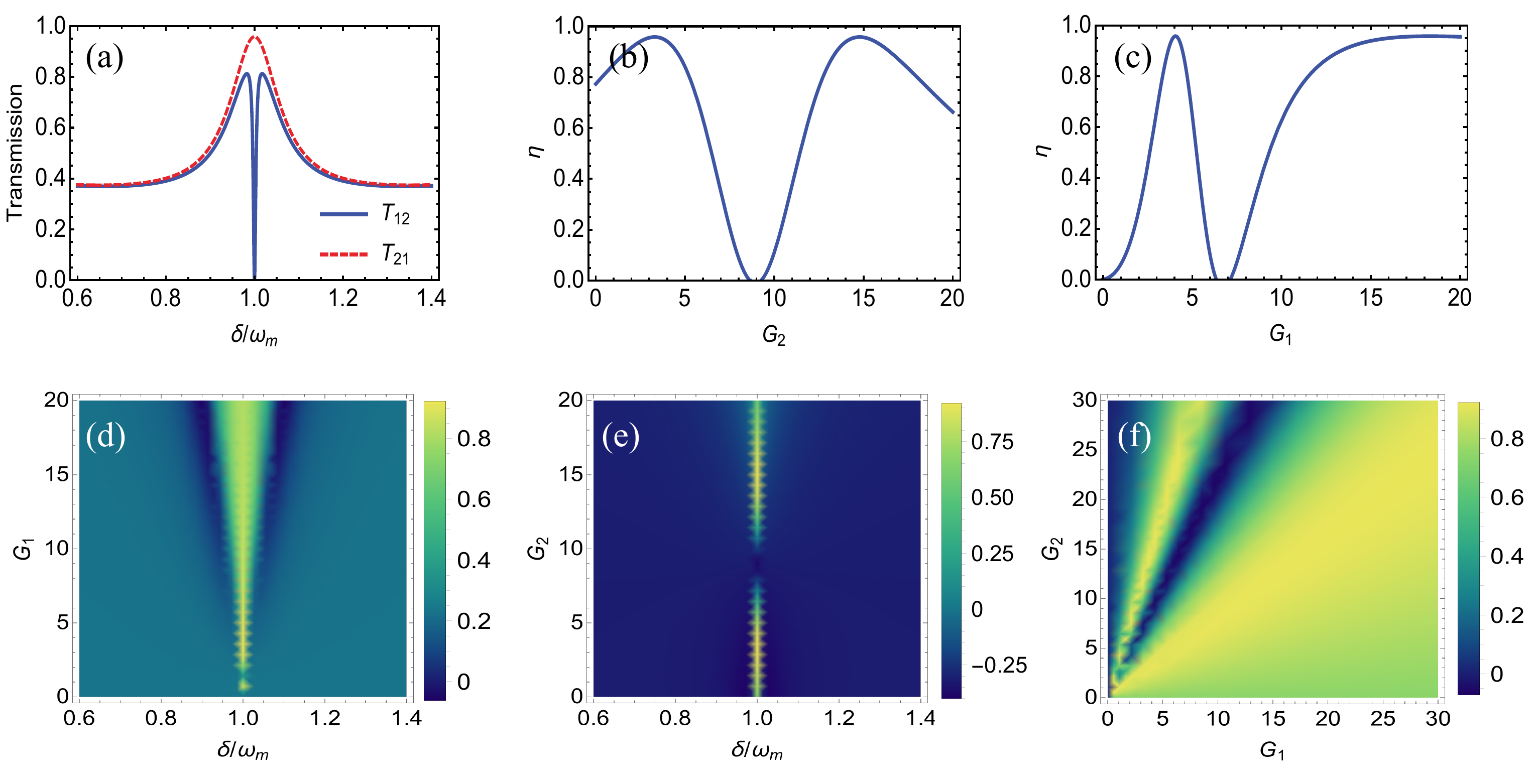}
    \caption{ Nonreciprocal transmission between $T_{12}$ (blue solid) and $T_{21}$(red dashed). (a) Transmission rate $T_{12}$ and $T_{21}$ as a function of detuning $\delta$ with $G_1=$ 4 MHz and $G_2=$ 15 MHz. (b) $\eta$ as a function of $G_2$ with with $G_1=$ 4 MHz. (c) $\eta$ as a function of $G_1$ with with $G_2=$ 3 MHz. (d) $\eta$ as a function of detuning $\delta$ and optomechanical strength $G_2$ with $G_1=$ 4 MHz. (e) $\eta$ as a function of detuning $\delta$ and optomechanical strength $G_1$ with $G_2=$ 3 MHz. (f) $\eta$ versus $G_1$ and $G_2$. The parameters are $\kappa_{01}=$ 2 MHz, $\kappa_{1,\alpha}=$ 100 MHz, $\kappa_{1,\beta}=$ 0.1 MHz, $\kappa_{0,2}=$ 200 MHz, $\kappa_{2,\alpha}=$ 500 MHz, $\kappa_{2,\beta}=$ 0.1 MHz, $J=$ 0.}\label{nonreciprocity}.  
    \end{center} 
\end{figure*}

In our system, there are several ways of interference: (1) the interference of the optical modes $a_1$ and $a_2$, which couples indirectly through the waveguide, can induce EIT-like line shape plotted with orange dashed line in Fig.~\ref{interference} (a), which was found in Ref.\cite{totsuka2007slow,xu2006experimental,xiao2009electromagnetically}, while in a waveguide-coupled microcavity system, the transmission spectrum is usually featured by symmetrical Lorentzian line shape as shown with the blue solid line. (2) the CW mode $a_{i,cw}$ couples with the mechanical mode by driving the pump field from CW direction. With changing the effective optomechanical coupling rate, we can always find the OMIT phenomenon in Fig.~\ref{interference} (b), which was studied in Ref.\cite{aspelmeyer2014cavity}. (3) the direct coupling of CW and CCW modes from each optical mode is another interference, resulting in mode-splitting by adjusting the photon-hopping interaction as shown in Fig.~\ref{interference} (c), and the interference between the CW and CCW modes was studied in Ref.\cite{mazzei2007controlled}. When all the interferences co-exist, the transmission becomes more complicated and we show the results in Fig.~\ref{interference} (d). With basic EIT-like shape, we consider the photon-hopping interaction $J$, and adjust the optomechanical coupling rate, the transmission spectrum shows OMIT (red dashed line) and OMIA (blue dashed line) phenomenon. The transmission rate can be controlled by choosing appropriate parameters. The parameters of red dashed line in Fig.~\ref{interference} (a) and (d) are $\kappa_{01}=$ 2 MHz, $\kappa_{1,\alpha}=$ 100 MHz, $\kappa_{0,2}=$ 200 MHz, $\kappa_{2,\alpha}=$ 500 MHz, $\kappa_{1,\beta}=\kappa_{2,\beta}=$ 1 MHz. The parameters of red dashed line in Fig.~\ref{interference} (b) and (c) are $\kappa_{01}=\kappa_{1,\alpha}=\kappa_{0,2}=\kappa_{2,\alpha}=$ 10 MHz, $\kappa_{1,\beta}=\kappa_{2,\beta}=$ 1 MHz. And the parameters of blue solid line in Fig.~\ref{interference} (a),(b) and (c) are $\kappa_{01}=\kappa_{1,\alpha}=\kappa_{0,2}=\kappa_{2,\alpha}=$ 10 MHz, $\kappa_{1,\beta}=\kappa_{2,\beta}=$ 1 MHz, $J$= 0, $G_1=G_2=$ 0. Other figures are given in the caption.

\subsection{Optical nonreciprocity and circulator}\label{sec3b}
Firstly, we study the case of weak interaction between CW and CCW mode where coupling strength $J$ is small, the $J_1$ and $J_2$ of the Eq. (\ref{eq9}) - (\ref{eq12}) can be omitted. We investigate the transmission between port 1 and port 2. According to the input-output relation
\begin{align}
    \varepsilon_{out} = \varepsilon_{in}-\sqrt{\kappa_{1,\alpha(\beta)}} a_{1,cw(ccw)}-\sqrt{\kappa_{2,\alpha(\beta)}} a_{2,cw(ccw)},
    \end{align}
and we denote $T_{mn} =|\frac{ \varepsilon_{out}^n}{\varepsilon_{in}^m}|^2$ as the transmission from port $m$ to port $n$. When the signal field is input on port 1 (i.e. $\varepsilon_s^{2,3,4}=$ 0),  
the transmission coefficient is 
    \begin{align}
        t_{12}=\frac{\varepsilon_{s}^1-\sqrt{\kappa_{1,\alpha}} a_{1,cw}-\sqrt{\kappa_{2,\alpha}} a_{2,cw}}{\varepsilon_{s}^1}.
    \end{align}

If the signal field is input on port 2 (i.e. $\varepsilon_s^{1,3,4}=$ 0), the transmission coefficient is 
    \begin{align}
        t_{21}=\frac{\varepsilon_{s}^2-\sqrt{\kappa_{1,\alpha}} a_{1,ccw}-\sqrt{\kappa_{2,\alpha}} a_{2,ccw}}{\varepsilon_{s}^2}.
    \end{align}
Substituting Eq. (\ref{eq9}) - (\ref{eq12}) into the transmission rate, one can rewrite the coefficients as
    \begin{align}
    t_{12} = 1-\frac{(\tau_1+\tau_2)\sqrt{\kappa_{1,\alpha}\kappa_{2,\alpha}}+\zeta_2\kappa_{1,\alpha}+\zeta_1\kappa_{2,\alpha}}{\zeta_1 \zeta_2 - \tau_1 \tau_2},
    \end{align}
    \begin{align}
    t_{21} = 1-\sqrt{\kappa_{1,\alpha}}\alpha_1-\sqrt{\kappa_{2,\alpha}}\alpha_2.
    \end{align}
The corresponding power transmission coefficient is given by $T_{12} = |t_{12}|^2$ and $T_{21} = |t_{21}|^2$.

We calculate the power transmission coefficient $T_{12}$ and $T_{21}$ versus $\delta$ which is shown in Fig.~\ref{nonreciprocity} (a). When the detuning $\delta=\omega_m$, transmission rate is $T_{12}=$ 0, but the oppsite one is $T_{21} \approx$ 1. It shows that the nonreciprocal optical transimission is enabled, where the signal can be transmit from port 2 to port 1, but the inverse process is forbidden. We define isolation coefficient $\eta=T_{21}-T_{12}$ to measure the degree of the nonreciprocity. Because the transmission is sensitive to the optomechanical coupling strength, we plot in Fig.~\ref{nonreciprocity} the $\eta$ as a function of $G_1$ and $G_2$ respectively at detuning $\delta=\omega_m$, from which $\eta \approx$ 1 for some values of $G_1$ and $G_2$, at which the nonreciprocal phenomenon is available. Fig.~\ref{nonreciprocity} (d) shows the isolation coefficient $\eta$ versus $\delta$ and $G_1$, in which high isolation rate can be achieved at $\delta=\omega_m$ with the increase of $G_1$. Similarly, by adjusting $G_2$, one can observe that the value of $\eta$ decrease when $G_2$ satisfies 3 MHz $<$ $G_2$ $<$ 9 MHz and increase when $G_2$ is 9 MHz $<$ $G_2$ $<$ 15 MHz. $\eta$ reaches its maximum value at the point of $G_2=$ 5 MHz and 19 MHz as shown in Fig.~\ref{nonreciprocity} (e). The effect of $G_1$ and $G_2$ on $\eta$  is asymmetric due to the difference of the loss rate of optical mode $a_1$ and $a_2$. Fig.~\ref{nonreciprocity} (f) illustrates how the isolation rate $\eta$ depends on $G_1$ and $G_2$, from which one can find that with the change of $G_1$ and $G_2$, the high isolation rate occupies a large proportion. The perfect nonreciprocity can be achieved by setting optical parameters. The physical mechanism of this phenomenon can be understood easily. Under the weak coupling of CW mode and CCW mode, there are two main types of interference involved in this system, i.e. indirectly coupling of the two optical modes and optomechanically coupling between CW mode and mechanical mode. Only CCW mode without optomechanically coupling is included in the transmission rate $T_{21}$ , which always results in EIT-like line shape. For transmission rate $T_{12}$, it contains two interference paths mentioned above and induces both OMIT and OMIA as shown in Fig.~\ref{interference} (d).

\begin{figure}[!ht]
    \begin{center}
    \includegraphics[width=8.5cm,angle=0]{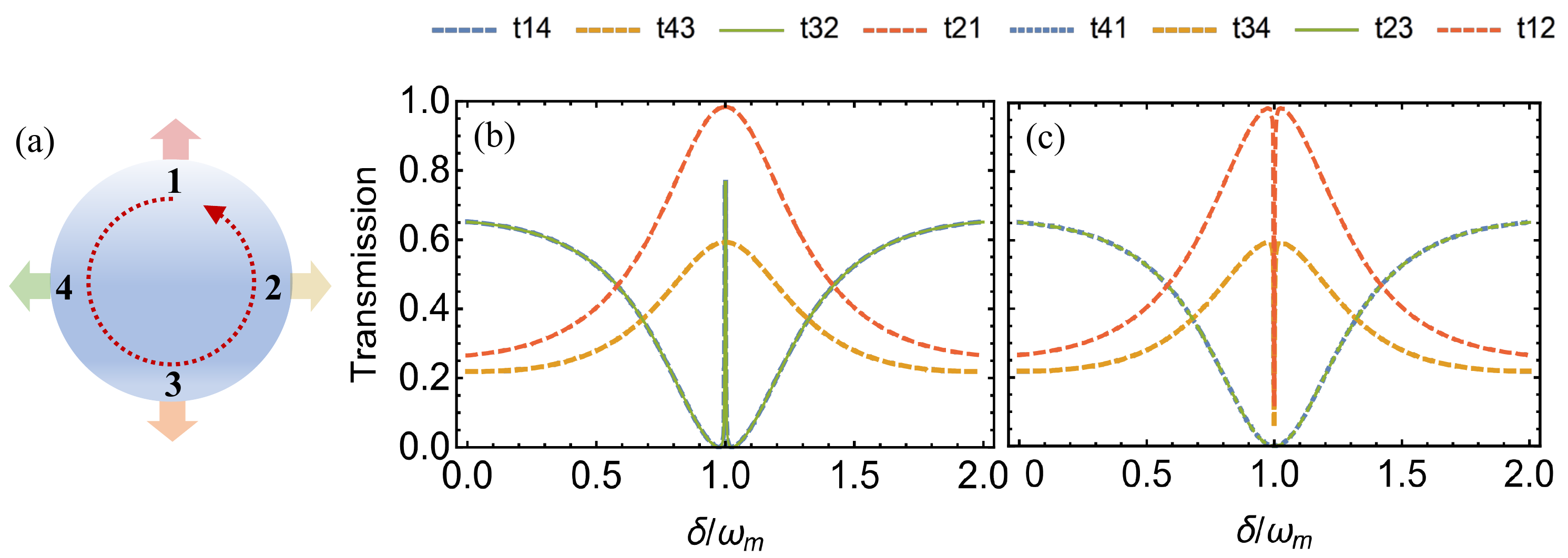}
    \caption{(a) Schematic of the circulator. (b) Transmission rate of $T_{x+1\rightarrow x}$.  (c) Transmission rate of $T_{x\rightarrow x+1}$. The parameters are $G_1=$ 5 MHz, $G_2=$ 1 MHz. $\kappa_{01}=$ 1 MHz, $\kappa_{1,\alpha}=$250 MHz, $\kappa_{1,\beta}=$ 0.01 MHz, $\kappa_{0,2}=$ 100 MHz, $\kappa_{2,\alpha}=$ 700 MHz, $\kappa_{2,\beta}=$ 800 MHz, $J=$ 0. }\label{circulator}
\end{center}
\end{figure}

Considering the situation of all four ports, because of the directional pump field, the coupling of CW mode and mechanic mode result in nonreciprocity, the transmission in port 1,2 is symmetric with that in port 3,4. When signal field input from port 1 (3), the time-reversal symmetry is broken and the system exhibits nonreciprocity. While signal field input from port 2 (4), the system functions as an add-drop filter. We define $T_{x\rightarrow x+1}$ as the signal transmission from the $x$-th to the $(x+1)$-th port and the reversal $T_{x+1\rightarrow x}$ for $x=$ {1,2,3,4}. In Fig.~\ref{circulator} (b), $T_{x+1\rightarrow x}$ is plotted as a function of detuning $\delta$, with relatively high transmission rate. And the transmission of opposite direction $T_{x\rightarrow x+1}$ is near-zero as shown in Fig.~\ref{circulator} (c). It means the signal is transferred from one port to  the adjacent port in a CCW direction ($4 \rightarrow 3 \rightarrow 2 \rightarrow 1$), but CW is forbidden as shown in Fig.~\ref{circulator} (a), which functions as a circulator.

\section{Optical routing}\label{sec4}

\subsection{Three-port routing}\label{sec4a}

\begin{figure}[!ht]
    \begin{center}
    \includegraphics[width=8.5cm,angle=0]{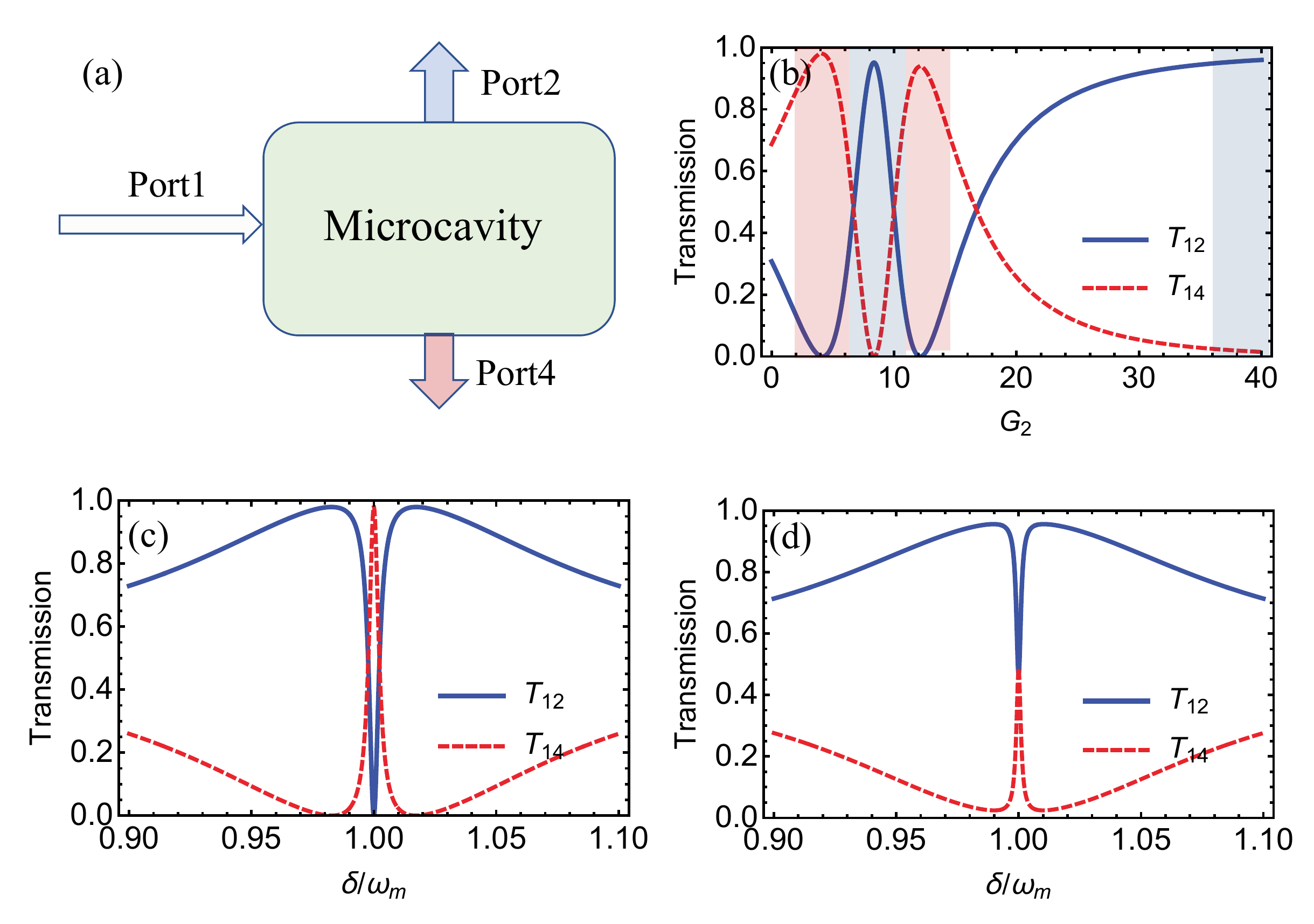}
    \caption{(a) Schematic diagram of a three-port optomechanical routing. (b) Transmission rates of $T_{12}$(blue solid) and $T_{14}$(red dashed) as function of $G_2$ with $J=$ 0. (c), (d) Transmission rate versus detuning $\delta$ with $G_2=$ 4 MHz, 10MHz. The parameters are $\kappa_{01}=$ 1 MHz, $\kappa_{1,\alpha}=$ 250 MHz, $\kappa_{1,\beta}=$ 1 MHz, $\kappa_{0,2}=$ 1 MHz, $\kappa_{2,\alpha}=$ 700 MHz, $\kappa_{2,\beta}=$ 200 MHz, $J=$ 0.}\label{router1}
    \end{center}
\end{figure}

\begin{figure*}[!ht]
    \begin{center}
    \includegraphics[width=16cm,angle=0]{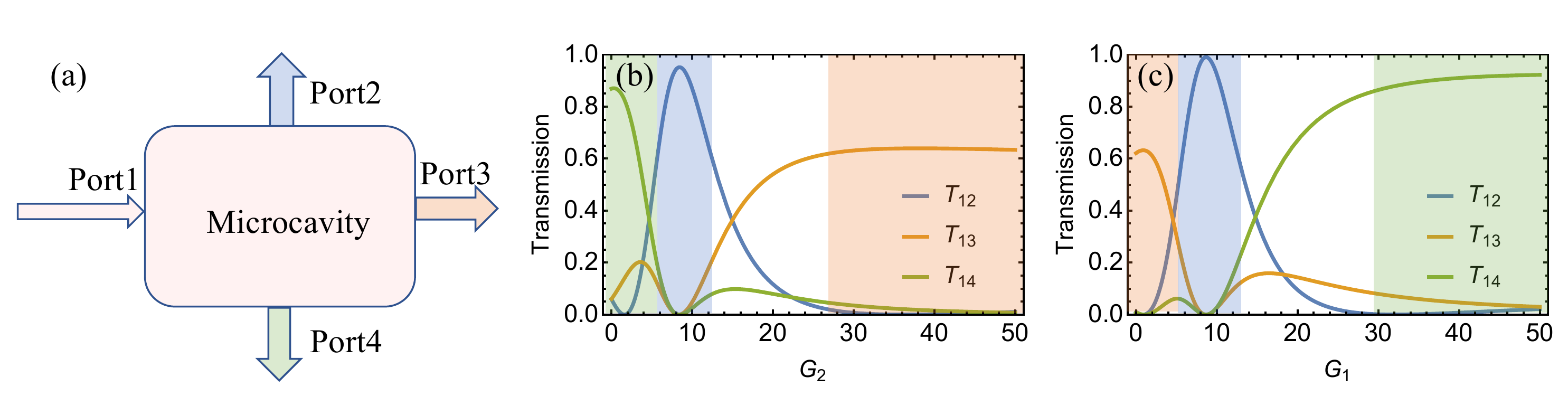}
    \caption{(a) Schematic diagram of a four-port optomechanical routing. Transmission rates of $T_{12}$ (blue), $T_{13}$(orange) and $T_{14}$ (green) are plotted as function of $G_2$ (b) and $G_1$(c). The parameters of (b) are $G_1=$ 5MHz $\kappa_{01}=$ 1 MHz, $\kappa_{1,\alpha}=$ 250 MHz, $\kappa_{1,\beta}=$ 1 MHz, $\kappa_{0,2}=$ 1 MHz, $\kappa_{2,\alpha}=$ 700 MHz, $\kappa_{2,\beta}=$ 400 MHz, $J_1$=$J_2=$ 50 MHz. And the parameters of (c) are $G_2=$ 10MHz, $\kappa_{01}=$ 1 MHz, $\kappa_{1,\alpha}=$ 150 MHz, $\kappa_{1,\beta}=$ 1 MHz, $\kappa_{0,2}=$ 1 MHz, $\kappa_{2,\alpha}=$ 200 MHz, $\kappa_{2,\beta}=$ 100 MHz, $J_1=$ 50 MHz, $J_2=$ 0.}\label{router2}
    \end{center}
\end{figure*}

Now we consider the signal field input from port 1, and calculate the transmission spectrum at port 2,3 and 4. The definition of $T_{12}$ is the same as before, and 
\begin{align}
    T_{13} = |\frac{-\sqrt{\kappa_{1,\beta}} a_{1,ccw}-\sqrt{\kappa_{2,\beta}} a_{2,ccw}}{\varepsilon_{s,1}}|^2,
\end{align}
\begin{align}
    T_{14} = |\frac{-\sqrt{\kappa_{1,\beta}} a_{1,cw}-\sqrt{\kappa_{2,\beta}} a_{2,cw}}{\varepsilon_{s,1}}|^2. 
\end{align}

We plot the numerical results in Fig.~\ref{router1} to show the possibility of routing implementation. For comparison, we first assume $J_1=J_2 =$ 0, indicating no scatting interaction. Output field from port 3 is absent in this case. The signal field input from port 1 can only output from port 2 and port 4 as shown in Fig.~\ref{router1} (a). Fig.~\ref{router1}(b) shows the transmission rate from port 2 and port 4 as a function of optomechanical rate $G_2$, from which we find that the output field of the two ports can achieve almost perfect routing. When $G_2 = 4$ MHz, the tranmission rates are $T_{13} = 1$ and $T_{12} = 0$ respectively, that is, the signal only outputs from port 2. With the increase of $G_2$, the signal strength of the two ports starts to convert including a balanced 50:50 beam splitter. When $G_2$ is increased to 8 MHz, the signal only outputs from port 4. We plot in Fig.~\ref{router1} (c)-(d) the transmission rate versus the detuning $\delta$ with $G_2 = 4$ MHz, 10 MHz, corresponding to a 100:0 and a 50:50 splitting rate, respectively. It can be shown that the three-port router can operate at any splitting ratio ranging from 0:100 to 100:0 with appropriate optomechanical coupling strength.

\subsection{Four-port routing}\label{sec4b}

When $J$ is large, the scattering effect is present, all three pathways are involved in the interference. The output field begins to appear on port 3. We show in Fig.~\ref{router2} how to route a photon in a deterministic way, i.e., photon inputs from port 1 can be routed to one of other three ports deterministically as shown in Fig.~\ref{router2} (a). Without loss of generality, we study the transmission rate at the point of detuning $\delta=\omega_m$. Since the system is sensitive to the optomechanical coupling strength, two different routing schemes are obtained by adjusting $G_1$ and $G_2$ respectively. In the first set of parameters, $G_2$ is adjusted as shown in Fig.~\ref{router2} (b), it can be seen that $T_{12}$ and $T_{13}$ can be almost suppressed while $T_{14}$ can be up to 90$\%$ around $G_2$ = 2 MHz. And if $G_2$ increases to 8 MHz,the signal can be routed to port 2 completely with efficiency close to 100$\%$. With the continued increase of $G_2$, port 2 and port 4 have almost no output, while port 3 has a nearly constant transmission rate of 0.6. Here, the transmission rate can not reach 1, mainly because  that the scattered light is the main component of port 3, photon is lost in the cavity or output from port 1. Different color region corresponds to different output ports in Fig.~\ref{router2} (b). It means, for an incident photon, target photon routing can be achieved deterministically by setting appropriate values of $G_2$. Under another set of parameters, we can achieve a similar function by adjusting $G_1$, except that the output ports are different when the optomechanical coupling coefficient is gradually increased as shown in Fig.~\ref{router2} (d). In this way, the multimode optomechanical system can be used as a controllable quantum router which the photon can emit from the desired target port with relatively high efficiency, which is attributed to the interference between different phonon excitation paths.

\section{Conclusion}\label{sec5}

In summary, we investigated the optomechanical multimode system that can be used to realize controllable optical nonreciprocity and routing, which is due to the multimode interference between different paths. We demonstrated that the nonreciprocal response is enabled when signal is input from opposite ports and nonreciprocity with high isolation can be realized by tuning the optomechanical coupling rates. Optical device such as diodes and isolators are possible because of the nonreciprocal phenomenon. We also showed that the system can also be used as  photon routing with three or four ports, in which photon can be transferred  from the input port to an arbitrarily target output port with high transmission. 

\section*{ACKNOWLEDGMENT}
The work was supported by the National Natural Science Foundation of China under Grants (11974205); National Key Research and Development Program of China (2017YFA0303700); Beijing Advanced Innovation Center for Future Chip (ICFC); The Key Research and Development Program of Guangdong province (2018B030325002).  
H.Z. acknowledges the China Postdoctoral Science Foundation under Grant No.2019M650620.
 M.W. acknowledges the China Postdoctoral Science Foundation under Grant No.2019M660605

\end{document}